\documentclass[a4paper,11pt,bibtotoc,liststotoc]{article}

\usepackage{amssymb}
\usepackage{graphicx}
\usepackage{hyperref}
\usepackage{harvard}
\usepackage{amsmath}
\usepackage[english]{babel}
\usepackage{amsfonts}
\usepackage{rotating}
\usepackage{pdflscape}
\usepackage{anysize}
\usepackage[official]{eurosym}
\usepackage{longtable}
\usepackage[a4paper,left=2cm,right=2cm,top=2cm,bottom=2cm]{geometry}
\usepackage{xspace}
\usepackage{array}
\usepackage{latexsym}
\usepackage{amsthm}

\newcommand{\ben}{\begin{enumerate}}
\newcommand{\een}{\end{enumerate}}
\newcommand{\bi}{\begin{itemize}}
\newcommand{\ei}{\end{itemize}}
\newcommand{\bd}{\begin{description}}
\newcommand{\ed}{\end{description}}

\usepackage{scrpage2}
\setheadwidth{textwithmarginpar}
\usepackage{setspace}%%
\usepackage{xspace}%% für besseres Handling der spaces
\usepackage{authblk}

\begin{document}

\begin{center}
\bigskip

{\LARGE On the plausibility of the latent ignorability assumption}

{\large \vspace{0.6 cm}}

{\Large Martin Huber} %\smallskip

\today

{\small {University of Fribourg, Dept.\ of Economics}}\\[0pt]
 %\textbf{\emph{preliminary and incomplete - comments very welcome}}
\end{center}

\noindent \textbf{Abstract:} {\small The estimation of the causal effect of an endogenous treatment based on an instrumental variable (IV) is often complicated by attrition, sample selection, or non-response in the outcome of interest. To tackle the latter problem, the latent ignorability (LI) assumption imposes that attrition/sample selection is independent of the outcome conditional on the treatment compliance type (i.e.\ how the treatment behaves as a function of the instrument), the instrument, and possibly further observed covariates. As a word of caution, this note formally discusses the strong behavioral implications of LI in rather standard IV models.  We also provide an empirical illustration based on the Job Corps experimental study, in which the sensitivity of the estimated program effect to LI and alternative assumptions about outcome attrition is investigated.
}

{\small \smallskip }


{\small \smallskip {\footnotesize Address for correspondence: Martin Huber, University
of Fribourg, Bd.\ de P\'{e}rolles 90, 1700 Fribourg, Switzerland, martin.huber@unifr.ch.}\thispagestyle{empty}\pagebreak }

{\small \renewcommand{\thefootnote}{\arabic{footnote}} %
\setcounter{footnote}{0}  \pagebreak \setcounter{footnote}{0} \pagebreak %
\setcounter{page}{1} }

\section{Introduction}

A frequently encountered complication when estimating the effect of a potentially endogenous treatment based on an instrumental variable (IV) methods is attrition/sample selection/non-response bias in the outcome. To account for this problem, the missing at random (MAR) assumption (e.g.\ \citeasnoun{Ru76b}), for instance, requires outcome attrition to only depend on observable variables. Alternatively, \citeasnoun{FrangakisRubin99} propose a latent ignorability (LI) restriction, which assumes attrition to be independent of the outcome conditional on the instrument and the treatment compliance type (i.e. whether one is a complier or non-complier in the notation of \citeasnoun{Angrist+96}). %\footnote{As the treatment is a deterministic function of the instrument and the compliance type, in addition explicitly conditioning on the treatment is redundant.}
In the IV framework both assumptions can be combined (e.g.\ \citeasnoun{MealliImbens04}), imposing independence conditional on the compliance type, the instrument, and further observables.

We argue that LI is nevertheless quite restrictive, as attrition is not allowed to be related to unobservables affecting the outcome in a very general way. Section \ref{sec2} formally discusses the strong behavioral implications of LI in standard IV models with non-response. This assumption should therefore be cautiously scrutinized in applications. As an example, consider \citeasnoun{Barnard+03}, who assess a randomized voucher program for private schooling with noncompliance (where the IV is the randomization and the treatment is private schooling) and attrition in the test score outcomes, because some children did not take the test.  %Indeed, one may argue that distribution of ability and motivation differs across compliance types: never takers less motivated able than compliers than always takers (check this with our bounds paper).
Unobservables as ability or motivation likely affect both test taking and test scores. LI (combined with MAR) requires that conditional on the compliance type (i.e.\ private schooling as a function of voucher receipt), voucher assignment, and observed covariates, test taking is not related to ability or motivation (and thus, test scores). Among compliers (only in private schooling when randomized in), those taking the test must thus have the same distribution of ability and motivation as those abstaining. However, even within compliers, heterogeneity in ability and motivation may be sufficiently high to selectively affect test taking such that LI fails.
Section \ref{sec3} provides an empirical illustration using the Job Corps experimental study, in which the estimated program effect under LI is compared to alternative assumptions about outcome attrition.

\section{IV models with nonresponse} \label{sec2}

Assume the following parametric IV model with nonresponse:
\begin{eqnarray}
Y=\alpha_0+D\alpha_1+ U,\textrm{ }\textrm{ } D=1(\beta_0+Z\beta_1  \geq V),\textrm{ }\textrm{ } R=1(\gamma_0+D\gamma_1 \geq W). \label{outcome}
\end{eqnarray}
$Y$ is the outcome of interest, $D$ is the binary (and potentially endogenous) treatment, and $R$ is the response indicator. Note that $1(\cdot)$ is the indicator function that is equal to one if its argument is satisfied and zero otherwise.  $Y$ is only observed if $R=1$ and unknown if $R=0$, implying non-response, sample selection, or attrition. $Z$ is a randomly assigned instrument affecting $D$ (but not directly $Y$ or $R$) and assumed to be binary, e.g., the randomization indicator in an experiment. $U,V,W$ denote arbitrarily associated unobservables, $\alpha_0,\alpha_1,\beta_0,\beta_1,\gamma_0,\gamma_1$ are coefficients.%which is equal to one if its argument is satisfied and zero otherwise.

\citeasnoun{Angrist+96} define four compliance types, denoted by $T$, based on how the potential treatment status depends on the instrument: An individual is a complier (defier) if her potential treatment state is one (zero) in the presence and zero (one) in the absence of the instrument and an always-taker (never-taker) if the potential treatment is always (never) one, independent of the instrument.  Assume that $\beta_1$ is positive (a symmetric case could be made for a negative $\beta_1$). Then, an individual is a complier if $\beta_0+\beta_1 \geq V>\beta_0$, an always taker if $\beta_0 \geq V$, and a never taker if $\beta_0+\beta_1<V$. Defiers do not exist due to the positive sign of $\beta_1$.

We now impose the following latent ignorability (LI) assumption, see \citeasnoun{FrangakisRubin99}, and critically assess it in the light of our standard IV model with attrition:\vspace{5 pt}\\
\textbf{Assumption 1 (latent ignorability):} $Y \bot R | Z, T$ (where `$\bot$' denotes independence),\vspace{5 pt}\\
which is equivalent to $Y \bot R | Z, D,T$ as $Z$ and $T$ perfectly determine $D$. Furthermore, we assume that the error term $U$ is continuous, such that $Y$ is continuous. Finally, for the moment we also impose that $U=V=W$ such that the same unobservable (e.g.\ motivation) affects the outcome (e.g.\ test score), treatment (e.g.\ private schooling), and response (e.g.\ test taking). %Considering the group of compliers, it can be shown that $U=V=W$ and Assumption 1 cannot hold jointly.
%$V$ burden and only if benefits outweigh burden we have participation.
%Need to hold that conditional on $R$, distribution of $U$ among compliers is the same across treatment states. E.g., for $R=1$

Note that Assumption 1 implies that the distribution of $U$ among compliers is the same across response states given the instrument:
%\begin{eqnarray}
%&&E(f(Y)| Z, T=c, R=1)=E(f(Y)|Z,  T=c, R=0)\label{check1}\\
%&\Leftrightarrow&E(f(U)| Z, T=c, R=1)=E(f(U)|Z,  T=c, R=0)\notag\\
%&\Leftrightarrow &E(f(U)| Z, \beta_0+\beta_1\geq U>\beta_0, \gamma_0+\gamma_1 \geq U) =E(f(U)| Z, \beta_0+\beta_1\geq U>\beta_0, \gamma_0 < U) \notag\\
%&\Leftrightarrow &E(f(U)| Z, \min(\beta_0+\beta_1, \gamma_0+\gamma_1) \geq U >\beta_0 )=E(f(U)| Z, \beta_0+\beta_1\geq U>\max(\beta_0, \gamma_0)), \notag
%\end{eqnarray}
%where $f(\cdot)$ denotes an arbitrary function with a finite expectation. This only holds if $
%\gamma_0+\gamma_1 \geq \beta_0+\beta_1$ and $\beta_0 \geq \gamma_0$, implying that all compliers are observed under treatment, as $R=1(\gamma_0+\gamma_1 \geq U)=1$ $\forall$ $U \leq \beta_0+\beta_1$, but none is observed under non-treatment, as $R=1(\gamma_0\geq U)=0$ $\forall$ $U > \beta_0$. Therefore, the treatment effect cannot be identified conditional on response due to a lack of variation in the treatment. Conversely, if there is treatment variation, Assumption 1 fails.
\begin{eqnarray}
&&E(f(Y)| Z=1, T=c, R=1)=E(f(Y)|Z=1,  T=c, R=0)\label{check1}\\
&\Leftrightarrow &E(f(U)| Z=1, \beta_0+\beta_1\geq U>\beta_0, \gamma_0+\gamma_1 \geq U) =E(f(U)| Z=1, \beta_0+\beta_1\geq U>\beta_0, \gamma_0+\gamma_1 < U) \notag, \notag
\end{eqnarray}
where $f(\cdot)$ denotes an arbitrary function with a finite expectation and the second line follows from the parametric model in \eqref{outcome}. Obviously, the joint satisfaction of $U=V=W$ and \eqref{check1} is impossible in this context, as the distribution of $U$ conditional on $\gamma_0+\gamma_1 \geq U$ and $\gamma_0+\gamma_1 < U$, respectively, is non-overlapping. An analogous impossibility result holds for $E(f(Y)| Z=0, T=c, R=1)=E(f(Y)|Z=0,  T=c, R=0)$, which is also implied by Assumption 1.

%Same problems carry over to mediation model with binary mediator:
%\begin{eqnarray*}
%Y&=&\alpha_0+D\alpha_1+\alpha_2M+U,\\
%D&=&1(\beta_0+Z\beta_1\geq V),\\
%M&=&1(\gamma_0+D\gamma_1 \geq W),\\
%\end{eqnarray*}
%Furthermore, consider the same model with continuous mediator: $M=1(\gamma_0+D\gamma_1 \geq W)$. LI is always violated, as both $M$ and $Y$ strictly increase in $U$ even within $\beta_0+\beta_1\geq V>\beta_0$.

Imposing $U=V=W$ seems too extreme for most applications and was chosen for illustrative purposes. However, even if the unobserved terms in the various equations are not the same, but non-negligibly correlated as commonly assumed in IV models, identification may seem questionable. Suppose, for instance, that $W=\delta_1 V + \epsilon$, where $\epsilon$ is random noise %(not correlated with $Z,V,U,W)$
and $\delta_1$ is a coefficient. Then, Assumption 1 and the model in \eqref{outcome} imply that
\begin{eqnarray}
&&E(f(U)| Z=1, \beta_0+\beta_1\geq V>\beta_0, \gamma_0 + \gamma_1 \geq \delta_1 V+ \epsilon) \label{corr}\\
&=&E(f(U)| Z=1, \beta_0+\beta_1\geq V>\beta_0, \gamma_0 + \gamma_1 < \delta_1 V+ \epsilon)\notag\\
&\Leftrightarrow& E\left(f(U)| Z=1, \min\left(\beta_0+\beta_1, \frac{\gamma_0+\gamma_1-\epsilon}{\delta_1}\right) \geq V >\beta_0 \right)\notag\\
&=&E\left(f(U)| Z=1, \beta_0+\beta_1\geq V>\max \left(\beta_0, \frac{\gamma_0+\gamma_1-\epsilon}{\delta_1}\right)\right). \notag
\end{eqnarray}
If $U$ is associated with either $\epsilon$, $V$, or both, the latter equality does not hold in general, but only if the association of $U, \epsilon$, $V$ is of a very specific form, which raises concerns about Assumption 1.

Finally, we investigate an in terms of functional form assumptions more general IV model, where $Y$, $D$, and $R$ are given by nonparametric functions denoted by $\phi$, $\psi$, and $\eta$, respectively:
%unobservables $U,V,W$  may now be vectors (rather than scalars) (is that true??) and arbitrarily associated with each other:
\begin{eqnarray}
Y=\phi(D,U),\textrm{ }\textrm{ }D=1(\psi(Z,V)\geq 0),\textrm{ }\textrm{ }R=1(\eta(D, W) \geq 0).\label{outcome2}
\end{eqnarray}
%Assume that $D$ is weakly monotonically increasing in $Z$ such that defiers are ruled out, which is one possible (and popular) approach to the identification of the local average treatment effect, see \citeasnoun{Imbens+94}. \citeasnoun{Vy02} shows that this allows representing $D$ without loss of generality by $D=1(g(Z)\geq h(V) )$, with $g,h$ being general functions. With a binary $Z$, the model simplifies to $1(\beta_0+Z\beta_1\geq h(V))$ which is similar to the parametric case. %However, the interpretation of the \beta_1 coefficient changes compared to: it now refers to an average rather than individual effect.
%Assume that similarly, $R$ is weakly monotonically increasing or decreasing in $D$, which appears plausible in several applications, see for instance \citeasnoun{Behagheletal2012}. Then, also the response model may be written similarly to the parametric case: $R=1(\gamma_0+D\gamma_1 \geq q(W))$, with $q$ denoting a general function. It follows that for $U=V=W$, the results in (\ref{check1}) carry over to the nonparametric model such that no common support exists in $D$ among observed compliers under Assumption 1. Furthermore, the identification issues related correlated errors, see (\ref{corr}), apply in a similar way.
%
%In our final IV model investigated, we relax monotonicity of response in the treatment (such that $1(\eta(D, W) \geq 0)$ does not simplify). In this case,
Under this model, Assumption 1 implies that
\begin{eqnarray}
E(f(U)| Z=1, \psi(1,V)\geq 0, \psi(0,V)< 0, \eta(1, W) \geq 0 )  =E(f(U)| Z=1, \psi(1,V)\geq 0, \psi(0,V)< 0, \eta(1, W) < 0).\label{corr2}
\end{eqnarray}
This can be satisfied in special cases, for instance if $U=\pi 1(\psi(1,V)\geq 0, \psi(0,V)< 0)+\varepsilon$, with $\pi$ denoting the (homogeneous) effect of being a complier and $\varepsilon$ being random noise.
%an unobserved factor which is independent of $Z, V, W$ and satisfies $E(\varepsilon|T)=0$, while $V$ may be arbitrarily associated with $W$.
Then, \eqref{corr2} simplifies to $E(f(\epsilon)| Z=1, \psi(1,V)\geq 0, \psi(0,V)< 0, \eta(1, W) \geq 0 )  =E(f(\epsilon)| Z=1, \psi(1,V)\geq 0, \psi(0,V)< 0, \eta(1, W) < 0)$, which holds because $\epsilon$ is independent of $W$. In general, identification requires that $T$ is a sufficient statistic to control for the endogeneity introduced by conditioning on $R$. This, however, implies that the association between $U$, $V$, and $W$ is quite specific, otherwise Assumption 1 does not hold.

%Assumption 1 requires that
%\begin{eqnarray}
%&&E(f(U)| Z, \beta_0+\beta_1\geq h(V)>\beta_0, \eta(1, W) \geq 0 )  =E(f(U)| Z, \beta_0+\beta_1\geq h(V)>\beta_0, \eta(0, W) < 0).\notag\\ \label{check2}
%\end{eqnarray}
%One can, of course, construct particular cases in which this assumption holds, for instance if the unobservable in the outcome equation is defined in the following way: $U=\pi 1(\beta_0+\beta_1\geq h(V)>\beta_0)+\varepsilon$, with $\pi$ denoting the (homogeneous) effect of being a complier and $\varepsilon$ being random noise, while $V$ may be arbitrarily associated with $W$.
%an unobserved factor which is independent of $Z, V, W$ and satisfies $E(\varepsilon|T)=0$, while $V$ may be arbitrarily associated with $W$.
%Then, (\ref{check2}) becomes
%\begin{eqnarray*}
%&&E(f(\epsilon)| Z, \beta_0+\beta_1\geq h(V)>\beta_0, \eta(1, W) \geq 0 )  =E(f(\epsilon)| Z, \beta_0+\beta_1\geq h(V)>\beta_0, \eta(0, W) < 0),
%\end{eqnarray*}
%which holds because $\epsilon$ is independent of $W$. More generally, however, (\ref{check2}) may well be violated, raising concerns about the plausibility of LI in empirical problems.

\section{Empirical illustration} \label{sec3}

As an illustration for treatment evaluation under LI and alternative assumptions about attrition, we consider the experimental evaluation of the U.S.\ Job Corps program (see for instance \citeasnoun{ScBuGl01}), providing training and education for young disadvantaged individuals. We aim at estimating the effect of program participation ($D$) in the first or second year after randomization into Job Corps ($Z$) on log weekly wages of females in the third year  ($Y$). Of the 4,765 females in the experimental sample with observed treatment status,
%\footnote{All in all, 5,805 females in the sample.}
wages are only observed for 3,682 individuals ($R=1$), while 1,083 do not report to work.

Reconsidering the IV model of (\ref{outcome2}), we assume that in each of $\phi$, $\psi$, and $\eta$ a vector of observed covariates, denoted by $X$, may enter as additional explanatory variables. Similar to \citeasnoun{FrHu14}, Section 2.2, we assume that (i) Assumption 1 holds conditional on $X$ (thus combining LI and MAR), (ii)  $U \bot Z | X, T$ such that the instrument affects the outcome only through the treatment, (iii) $T \bot Z| X$ which is implied by random assignment, (iv) $\Pr(T=c)>0$ and $\Pr(T=d)=0$ so that compliers exist and defiers are ruled out, and (v) $0<\Pr(Z=1|X)<1$, ensuring common support in the covariates across instrument states.  $X$ (measured prior to randomization) includes education, ethnicity, age and its square, school and working status, and receipt of Aid to Families with Dependent Children (AFDC) and food stamps.

%Reconsidering the IV model of (\ref{outcome2}), we assume that in each of $\phi$, $\psi$, and $\eta$ a vector of observed covariates, denoted by $X$, may enter as additional explanatory variables. We impose comparable assumptions as \citeasnoun{FrHu14} (for their first outcome period):\vspace{5 pt}\\
%\textbf{Assumption 1a (latent ignorability combined with MAR):} $Y \bot R | X, Z, T$,\\
%\textbf{Assumption 2 (exclusion restriction):} $U \bot Z | X, T$ and $T \bot Z| X$, \\
%\textbf{Assumption 3 (Monotonicity):} $\Pr(T=c)>0$, $\Pr(T=d)=0$, and $0<\Pr(Z=1|X)<1$. \vspace{5 pt}\\
%Assumption 1a weakens Assumption 1 to hold conditional on $X$, combining LI and MAR. Here, $X$ (measured prior to randomization) includes education, ethnicity, age and its square, school and working status, and receipt of Aid to Families with Dependent Children (AFDC) and food stamps (proxying social status). Assumption 2 imposes that conditional on the type and observed covariates, the instrument is related to the outcome only through the treatment. Furthermore, the instrument must be independent of the types given the covariates, which holds under random instrument assignment. Assumption 3 requires that compliers exist, defiers do not exist, and that there is common support in the covariates across instrument states.
\begin{table}[ht]
\caption{Descriptive statistics}
\label{descr}
\begin{center}
{\scriptsize
\begin{tabular}{r|cc|cc|cc}
  \hline\hline
  & \multicolumn{2}{c|}{total sample}  &
\multicolumn{2}{c|}{working} & \multicolumn{2}{c}{not working} \\
  & mean & std.dev & mean & std.dev & mean & std.dev \\
  \hline
 education: 12 years   & 0.23 & 0.42 & 0.25 & 0.44 & 0.17 & 0.37 \\
  education: 13 or more years & 0.03 & 0.18 & 0.04 & 0.19 & 0.01 & 0.10 \\
  race: black & 0.54 & 0.50 & 0.53 & 0.50 & 0.56 & 0.50 \\
  race: Hispanic & 0.19 & 0.39 & 0.18 & 0.38 & 0.21 & 0.40 \\
  age & 18.59 & 2.18 & 18.66 & 2.19 & 18.37 & 2.14 \\
  in school prior to randomization & 0.63 & 0.48 & 0.63 & 0.48 & 0.61 & 0.49 \\
  school information missing& 0.02 & 0.14 & 0.02 & 0.13 & 0.03 & 0.17 \\
  in job prior to randomization & 0.61 & 0.49 & 0.65 & 0.48 & 0.47 & 0.50 \\
  received AFDC & 0.41 & 0.49 & 0.40 & 0.49 & 0.45 & 0.50 \\
  received food stamps & 0.54 & 0.50 & 0.52 & 0.50 & 0.60 & 0.49 \\
  treatment: Job Corps participation & 0.45 & 0.50 & 0.46 & 0.50 & 0.41 & 0.49 \\
  instrument: randomization & 0.64 & 0.48 & 0.66 & 0.48 & 0.60 & 0.49 \\
  instrument: kids under 6 & 0.77 & 0.90 & 0.73 & 0.88 & 0.88 & 0.95 \\
  instrument kids under 15 & 1.15 & 1.26 & 1.12 & 1.23 & 1.25 & 1.34 \\
  \hline
\end{tabular}
}
\end{center}
\end{table}

We compare sempiparametric LATE estimation based on the latter assumptions (see Theorem 1 in \citeasnoun{FrHu14}) to (i) MAR-based LATE estimation as in Section 2.3 of \citeasnoun{FrHu14} (assumptions: $Y \bot R | X, Z, D$, $(U,T) \bot Z | X$,  $\Pr(T=c)>0$, $\Pr(T=d)=0$, $0<\Pr(Z=1|X)<1$), (ii) the Wald estimator among those with $R=1$ (ignoring sample selection), and (iii) the method of \citeasnoun{FrFrHuLe2015}, which tackles sample selection and treatment endogeneity by two distinct instruments. In the latter approach, which allows for non-ignorable selection related to $U$ in a more general way than LI, we use the number of kids younger than 6 in the household 2.5 years after random assignment as instrument for $R$. We apply a semiparametric version of the estimator outlined in equation (23) of \citeasnoun{FrFrHuLe2015} along with the weighting function in their expression (21). %, see \citeasnoun{HuberMellace2011} for a discussion and a test of the IV validity of number of kids.

\begin{table}[ht]
\caption{Effect estimates }
\label{results}
\begin{center}
{\scriptsize
\begin{tabular}{r|cccc}
  \hline\hline
  & LI + MAR &  MAR & Wald &  2 IVs \\
  \hline
effect &  0.12 & 0.16 & 0.12 & 0.16 \\
 standard error &  0.06  & 0.06 & 0.05 &  0.33  \\
  bootstrap p-values (quantile-based) &  0.05 & 0.00 & 0.03 & 0.65  \\
   \hline
\end{tabular}
}
\end{center}
\end{table}

%In the latter approach, which allows for non-ignorable selection that is related to the unobservable $U$ in a more general way than LI, we use the number of kids younger than 15 in the household 2.5 years after random assignment as instrument for $R$. We apply a semiparametric version of the estimator outlined in equation (23) of \citeasnoun{FrFrHuLe2015} along with the weighting function in their expression (22). %, see \citeasnoun{HuberMellace2011} for a discussion and a test of the IV validity of number of kids.

%\begin{table}[ht]
%\caption{Effect estimates }
%\label{results}
%\begin{center}
%{\scriptsize
%\begin{tabular}{r|cccc}
%  \hline\hline
%  & LI + MAR &  MAR & Wald &  2 IVs \\
%  \hline
%effect &  0.12 & 0.16 & 0.12 & 0.09 \\
% standard error &  0.06  & 0.06 & 0.05 &  0.61  \\
%  bootstrap p-values (quantile-based) &  0.05 & 0.00 & 0.03 & 0.69  \\
%   \hline
%\end{tabular}
%}
%\end{center}
%\end{table}

Table \ref{descr} provides descriptive statistics for the covariates, the treatment, and the instruments in the total sample and for working and not working females.  Across the latter groups for instance education, aid receipt, previous job status, and Job Corps participation differ importantly, pointing to non-random selection into employment. Table \ref{results} presents the effect estimates, standard errors, and p-values based on 1999 bootstraps using the quantile method. The effect under LI + MAR (based on Theorem 1 of \citeasnoun{FrHu14}) of 0.12 log points virtually identical to the Wald estimator which ignores sample selection bias, and both are statistically significantly different from zero. The MAR-based estimate is one third higher, but not significantly differently so. The method of \citeasnoun{FrFrHuLe2015} based on two instruments (2 IVs) yields virtually the same effect as MAR and is neither statistically significantly different from any other estimator, nor from zero at any conventional level. %The method of \citeasnoun{FrFrHuLe2015} based on two instruments (2 IVs) yields a somewhat smaller wage effect than LI+MAR and the Wald estimator, but is neither statistically significantly different from the other estimates, nor from zero.

It seems important to understand the differences in the behavioral assumptions of the estimators. LI + MAR, for instance, assumes that given the covariates and program assignment, unobservables like ability and motivation do not jointly affect employment and wages among compliers. In constrast, the method of \citeasnoun{FrFrHuLe2015} does not rely on this restriction and allows for more general forms of sample selection, at the cost of also requiring a valid instrument for employment. In our illustration, the results turned out to be rather robust to the different assumptions considered, which need not necessarily hold in other contexts.

% KOMMENTARE VON MARKUS:
% Latent ignorability funktioniert eigentlich nur, wenn Compliance Typ eine sufficient statistic für die unbeobachtbaren Variablen sind, also die relevanten unbeobachtbaren Variablen sich in einer Drei-Punkt-Verteilung zusammenfassen lassen.
%Momentan liest sich das Papier eher so, als ob wir eine Kritik an unserem eigenen JASA Papier üben und stattdessen den Heckman Schätzer empfehlen.
%Eine Option wäre nun zu warten, bis wir unser Attrition Papier mit den ZWEI Instrumenten geschrieben haben. Das wäre dann nämlich ja auch die passende Alternative zum Heckman Schätzer, oder ?  Dann könnten wir nämlich argumentieren, daß Latent Ignorability eine durchaus strenge Annahme ist und daß es besser wäre, unseren neuen Schätzer mit ZWEI Instrumenten zu nehmen (statt den Heckman Schätzer zu empfehlen).

\setlength\baselineskip{14.0pt}
\bibliographystyle{econometrica}
\scriptsize{
\bibliography{Froelich}
}
\end{document}